# Estimation of Gravitational Constant *'$G_N$'* from CMBR data and Hubble's parameter

**G.SURYAN, Physics Dept., Indian Institute of Science, Bangalore, 560012, India.**

In the companion submission by the author entitled "What is Time ? A New Mathematico-Physical and Information Theoretic Approach" broad outlines of the Nature of Time that could be formulated on the basis of a finite coarse grained and growing universe named CHALACHALA with the most basic entities called EKONS has been sketched. Planck's Quantum of action has been given in the form of angle (ratio). It occurred to the author that once the basic entities are known other fundamental physical constants like *$G_N$* may also be amenable for estimation through observational data particularly information regarding expansion of universe through Hubble's parameter. In order to proceed further it is necessary to highlight some of the key features of the procedure as follows.

Depending upon the number of Ekons agglomerating one can envisage the formation of various particles and radiation of the universe. Their stability would imply that the degrees of freedom of the constituent Ekons would be effectively frozen corresponding to a degenerate assembly. This would also apply to the so called cold dark matter / radiation with very large effective 'Fermi level' (like those of metals in solid state physics) with a steep drop ( narrow width) in occupation number at the Fermi level and its own effective temperature. This width would give the thermodynamic temperature of the so called cold dark matter. With Time this width would also grow and the corresponding effective mass would also be very small. Once the effective mass is estimated it would be possible to estimate the 'cold' dark matter - consisting of mostly the individual "EKONS" not otherwise agglomerated in to matter or radiation quantum.

These have been put pictorially in (Fig. 1). It may be noted here that the distribution is highly degenerate and the fall in the occupation number is very steep and direct numerical differentiation is not possible. Fortunately this aspect has already been tackled in solid state theory of metals and semi conductors. This is done by noting that a differential coefficient of the Fermi function for one order can be related to that of a neighboring order with very good accuracy. Procedures and tables are available from the excellent compilation given by Blakemore[2] . The Fermi level is an estimate of the size of the universe. The expansion is related to the Hubble constant .

Attempts to introduce the quantum of action in general relativity have been singularly unsuccessful except for the work of Hawking who found that not only a massive star but also a ultra mini space-time singularity collapsing under its own gravitational attraction does have influence outside and related to the quantum of action.

Even if one EKON moves with respect to the rest of the universe there has to be a reaction. Hence, for momentum conservation the momenta has to be in the ratio $N_W$ :1. Therefore the <u>effective mass</u> $m_e$ of one EKON is $1/N_W$ of the mass of the universe. $N_w$ is known from the size of the universe and the extent of a single EKON has been assigned $[(h/2\pi)/2]^3$ and the space part is $[(h/2\pi)/2]^{3/2}$.

For getting the momentum þ & wavelength associated with assembly of Ekons constituting either particles or quanta we can use the existing super structure of Quantum Mechanics through the relation $m^* = þ^{2\lambda}/2hc$, $þ^2 \equiv þ_x^2 + þ_y^2 + þ_z^2$.

The effective mass $m_e$ of a single Ekon can be estimated by knowing the number of Ekons constituting the massive particle or quanta of radiation. For estimating the number of Ekons giving raise to a massive stable particle would be very high and one can get precise values of the large numbers through the properties of prime numbers.

**Role of Prime Numbers**

The number of EKONS constituting the observed particles like electron, proton, neutron and the alpha particles are very large. They are highly stable. If the numbers of EKONS constituting these entities were not prime they can break up into smaller entities corresponding to the prime number factors, impairing their stability. Hence there is great need to really understand large prime numbers their squares and their reciprocals. For useful specific properties and distribution of Prime Numbers we can refer to Ribenboim[3]. In the observation and description of the universe we live in, all windows available to us are of electromagnetic origin from microwave to the γ rays.

We have yet to tap other non-electromagnetic forms like gravitational waves having their own temperature. This form may be identified with cold dark matter far larger in number than baryonic matter and electromagnetic radiation. Joule in his experiments conclusively demonstrated that mechanical work is converted to heat and increasing temperature. Left alone this heat gets converted to radiation and decrease of temperature. The temperature and entropy concepts are closely related to statistical mechanics with its own fundamental physical constant $k_B$ . Thermodynamics introduces absolute temperature scale and T.

For attaining thermodynamic equilibrium not only large numbers but also rapid interaction are essential. These are fully provided for through the strong repulsive interactions and through gravitational forces in the long range as well as net generation. In Mother Nature the basic entities tend to conform to one another and they become identical with the result that they only survive. This conformity enables one to have the information in a concise form using the permutation symmetry. The information content is expressible in the form of a logarithm and that information itself in log-log form and so on. As already remarked the time scale of such interaction is set by the random *net* generation of EKONS. The emphasis on *net* is intentional. An estimate of the current net generation may be obtained through the time rate of change of the Hubble's parameter.

Apart from the accurate temperature and low anisotropy the most important reliable result from CMBR observations is that the universe is flat as per the multi-moment analysis recently reported by p.de Bernardis[4] et al. Quantitatively this means that the following relation holds $\rho_{(critical)} = 3H_o^2/8\pi \cdot G_N$

This may be inverted to give $G_N = (3H_o^2/8\pi)1/\rho_{critical}$

**Estimating $\rho_{critical}$**

We can relate the total number of EKONS by the Wheeler number $N_W$ already introduced and the 'spatial' part of the volume is $[(h/2\pi)/2]^{3/2}$. From the known volume of the universe and the total number of the Ekons we can get the mass density.
One may wonder whether the value of $\rho_{critical}$ has not been calculated by assuming the value of $G_N$. In the present model of a growing universe it is postulated that effective mass density is just critical and is a result of the basic repulsive character (non overlapping) and the net generation of Ekons, which is also a means of reckoning and ensuring the unidirectionality of Time. In order to compete this program an estimate of the effective mass of an individual Ekon or the effective mass of a known assembly of Ekons constituting a well known particle or quantum of radiation.

Analysis of the CMBR data gives a good value for the temperature combined with the Boltzman's constant can give the energy and the wave length of the photons. This may be used to get an effective momentum for the photons and evaluate the contribution at CMBR to the over all matter/energy density of the universe due to electromagnetic radiation which of course is very small. However one can consider waves of the size of the universe from which one can get idea of the effective wave length. In this approach one is thinking in terms of a standing wave, and this can be converted into an estimate of the effective mass of a group of Ekons.

## Velocity of light

The velocity of light has been playing a leading role in all physics including cosmology and recent efforts like M-branes and so on. Currently the velocity of light is just assigned by relating it to the electric and magnetic permeability $\varepsilon_o, \mu_o$, through the relation $c^2 = 1/\varepsilon_o\mu_o$. This can permit both $\varepsilon_o$ and $\mu_o$ to be negative or for that matter have complex conjugate exponential factors without affecting the numerical results.

**Inertia and Energy concepts :** Having taken the position that inertia has to be understood as an expression of the Mach's principle and that the entire universe has to react , the question will arise as to the meaning of Energy and the meaning of relation $E = mc^2$. All forms of energy have their mass equivalent.

There have been many attempted derivations of the mass energy relation as thoroughly researched by Max Jammer[5] who concludes that the nature of the mass energy relation is still an open question. Author has pertinent views on the role of the velocity of light and the mass energy relationship.
In order to link the authors approach to the existing super structure the mass energy relation is used with the proviso that it is better to use momentum rather than energy for numerical estimates where appropriate.

Ekons have been assigned the value of $[ ½ (h/2\pi )]^3$ units being the product of both spatial and momentum parts. Spatial part can be identified with potential energy and the momentum to Kinetic energy. Assuming the virial theorem that the average of kinetic energy may be equated to the average potential energy, one may take the square root of the EKON as volume, the space part "Potential" and the momentum part "Kinetic".
For this purpose it is best to use $E = p^2/2m*$ as an expression for energy and allow **m** to take positive, negative or even complex values. Such an approach is already prevalent in solid state physics.

It must be remembered that only differences of energy are physically meaningful.
In the quantum regime energy is related to time through a frequency $\nu$ or wave length $\lambda$ .

The EKONS have no mass but acquire inertial property by the reaction of the entire universe (i.e by the application of the Mach's principle enunciated by Mach more than a century ago). It is necessary to have an estimate of the effective mass. In fact the position taken by the present author is to give full credence to Mach's principle as illustrated in the companion submission "What is Time? A New Mathematico- Physical and Information Theoretic Approach crucial step.

i) Associate a wave length with Ekon by taking the "Space" part of it. Say it is $\lambda$ and using the well known relation $\lambda = (h/m_\varepsilon ) v$ , $m_\varepsilon$ = effective mass of a single Ekon. With these the number of Ekons constituting the electron has been estimated and are listed in" Numerical Estimate I". The number of EKONS constituting the proton , the neutron and the alpha particle can be similarly estimated from the known mass ratios.

## Links to thermodynamics and statistical mechanics

Contrary to the currently favored big bang theory, the present model proposes that the entities are created and annihilated randomly and the net go on to form various
physical entities like radiation quanta and material particles with their own temperatures and disorder increases with time. It is therefore possible to estimate the gravitational constant $G_N$ from the observed thermal parameters from CMBR data and Hubble parameters.

It must be emphasized that if the time parameter "t" is reversed in sign then the gauge invariance follows and the rest of electromagnetism and its consequences stay in place.

In systems however big, if there is an overall recurrence period then systems can come to their original configuration exactly like what is envisaged in Poincare cycles.

Maxwell's equations in their linear form are invariant with respect to Lorentz transformation and hence Einstein gave time the status of an additional dimension through the velocity of light.

However, if Maxwell's equation are rewritten along with source terms then they will no longer be linear and that it is necessary to do so has been emphasized by many. For details refer to Lakhtakia[6].

Following points would be in order : The assigned compound "volume" for a single EKON is $[1/2 (h/2\pi)]^3$ therefore the spatial part is and $[1/2 (h/2\pi)]^{3/2}$ the momentum part is also $[1/2 (h/2\pi)]^{3/2}$. One can choose the potential outside as zero. This prescription is applicable at all finite scales of calculation. There is no need to bring in the infinity. Only differences matter as far as energy is concerned. It is well known that conventional methods for evaluating the Planck's Constant have been made by using either h/e or h/k or $h\nu = E_1 - E_2$.

It should be noted that in the finite model chosen the universe as a whole will not have a net gravitational field extending out. There will be a fall off. The only way gravitation can extend itself will be due to growth by net generation. This may appear strange and not easily grasped by people too deeply committed to the concept of an infinite space and time. Actually the generation and repulsive interaction leading to expansion just manage to keep the universe from either shrinking back and whither away or expanding very rapidly; i.e a closed but expanding universe. Recent experimental evidence in the form of elaborate multipole expansions of the CMBR radiation as reported by Bernardis et. al is highly significant and is sufficiently conclusive evidence for a just flat universe.

**The numerical calculations.** These have been summarized in the accompanying "Numerical Estimates II & III ".

It is important to recall the work of Willard Gibbs[7] on the relationship between statistical mechanics and thermodynamics almost a century ago. In the model proposed EKONS (all alike) constitute essentially a micro-canonical equilibrium. It must be noted that the entire assembly is essentially cold.

The random generation of new EKONS provides the necessary input to keep the universe quasi stationary. This is ably summarized by Eddington "As regards heat energy, Temperature is a measure of its degree of organization. The lower the temperature the greater the disorganization! There is safety in numbers".

**Over view of the results of Ekons/Chalachala model**

1. The $\Lambda$ problem ; that is the $\Lambda$ parameter is $\approx 10^{-120} M_{pl}^2$  $M_{pl}$ = Planck Mass is a factor of $10^{-60}$ ratio. The characteristic effective mass of the Ekon is about $5.67 \times 10^{-52}$ kg vs Planck Mass $2.1767 \times 10^{-8}$ kg is 40 orders of magnitude smaller.

2. Very few if any of blue shift as compared to Red shift. This indicates that the major entities of the universe like galaxies are all moving away from one another. This is easily understood in the new model proposed now by noting that the inter galactic space is expanding with time on a large scale. New entities created repel one other.

3. When created the part of the new entities constitute matter (including radiation) gets pushed in between the existing entities.

4. It is important to note that time scales govern the mathematical representation of most phenomena. Not only the scales but also the true Time's arrow.
One may imagine that new creation takes place and expands the universe by expanding say an atom or a proton or an electron. It is not so. The enormous stability and identity of fundamental particles rule out such a possibility.  The apparent smoothness and continuity of space and time is beguiling but the fundamental granularity cannot be wished away.

5. Thermodynamic Arrow of Time
A best statement about the unique supremacy of the thermodynamic statistical mechanics made by Eddington is as relevant today as seventy five years ago and the same has been remarked in our previous "**What is Time? A New Mathematico-Physical and Information Theoretic Approach.**

6.The procedure adopted to estimate the chalachala constituents in terms of a number of Ekons and their density is as follows :

a) The volume of the universe has been estimated from the available data as given above. It is assumed that the universe is steadily expanding as a result of the continuous net generation of ekons . The volume of the universe has been estimated from the present day Hubbles constant and the age of the universe is taken nominally $10^{10}$ years which is equivalent to $3.155815 \times 10^{17}$ sec. in order to calculate the number of ekons $[1/2( h/2\pi )]^{3/2}$ is taken and this enables us to calculate total number of ekons as $8.865979 \times 10^{116}$ which gives equivalent of the number density of ekons as $3.315661182 \times 10^{66}$ per m$^3$. From the estimated mass of the universe one can get $m_\in$ Ekon mass = $5.67 \times 10^{-52}$ kg. If one compares this with the $m_{pl}$ the ratio comes to be <u>$2.6049 \times 10^{-75}$</u>. This shows that the true mass content of the universe is determined by the very large number of ekons . This clearly solves the so called cosmic co-incidence problem giving some credence to the ekon concept utilized.

One of the well known mass figure of stable particles namely the electron, the estimate comes to be $1.266608 \times 10^{47}$. It is well known that electron is one of most stable entity in the universe and therefore the number of ekons constituting the electron must be a prime number. If it is not a prime number it would be possible for some type of standing wave structure developing and affect the stability of the particle.

The only method of avoiding such a standing wave is provided by letting a number of ekons per electron to be a prime number. There are various of algorithms for finding out prime numbers in the vicinity of a known number, thereby we can find the number of ekons per electron and estimate the mass of a single ekon . Similarly it should be possible to estimate the number of ekons constituting a proton and through them the specific figures for various other particles like the neutron, deuteron and other exotic particles like quarks.

Making use of the known critical density it is possible to estimate a value of the gravitational constant $G_N$. The results are given in "Numerical Estimates I".

7. One of the challenges posed by deep analysis by several authors is to explain how the entropy is very low at the beginning . In this connection one can give examples of "cold" items where the energy is stored with an effective zero entropy. Following are the examples.

Examples of "Cold" energy stored
a) Rotating fly wheel
b) Energized super conducting magnet
c) A mass separated from a much larger mass
d) A charged electric cell

Note that the mutual conversion of the above forms of energy are essentially reversible and they are all "cold" with very low entropy. Transfer of energy and momentum on collision and conservation of both cannot be satisfied simultaneously. Example Compton effect in a free atom. In dealing with ekons it must be remembered that it is most important to deal with momentum rather than energy. Conservation of momentum and angular momentum is very essential. Hence the main issues posed an understanding cosmology have been tackled by chalachala concepts and gives credence to a presence of "cold dark entities".

Analysis of the CMBR experimental results gives a precise value for the temperature. This enables us to get the estimate of the number of ekons which can constitute a photon.

8. In summary the present approach has shown time as to be treated as unidirectional and how it fits in neatly with the thermodynamics arrow.

9. Some of the finer points of the Ekon approach will be clear when one is dealing with photons. Of the various cosmological observations the most precise ones are regarding CMBR by which we can get thermodynamic temperature of the photons, particularly the peak around 160 mega hertz because the photons are already moving with the velocity of light. The photons rest mass would be zero. Therefore one has to deal with the momentum of the photons rather than anything else.

This calculation has been given in the accompanying table "Numerical Estimates III. Here we are dealing with the momentum part and one can get the space part in terms of wave length which can be related to the number of ekons needed to get a observable photon. It must be remembered that a single ekon totally unobservable and it requires at least 16 entities to cover the whole photon which consists of alternate Potential and Kinetic energy.

In the Einstein formulation of relativity it should be noted that the velocity of light is taken as fundamental and to confirm to the Lorntz transformation, the line element Time is used. It is quite efficient as a velocity and squared (light) by making the line element in the form of a square of a length. It must be noted that the sign is lost and cannot be distinguished between the past and the present. Some time the quantity $\sqrt{-1}$ is used in order to take care of the convention in the line element. Similarly $C^2$ is used in converting Reiman Christophel tensor to Einstein tensor $G\mu\nu$ and a $\Lambda$ term for growth. The new formulation in terms of ekons and their assemblies Chalachala gives a better understanding of cosmology.

10. One of the most important requirement for any new theory about the foundations of physics is to ensure that a existing super structure is preserved to the extent possible. One such requirement, a major one at that, is the limitation of the speed of light and relationship between energy, mass and speed of light square.

We have enunciated that mass is due to reaction of the entire universe to any changes anywhere. Such changes can take place by ekons turning from what one may call from potential to kinetic and kinetic to potential and so on. For this to happen it is necessary that adequate supply of small groups of ekons are single ones and are available in the neighborhood. In all this what is important is to remember what one traditionally calls the conservation of momentum and orbital angular momentum and last but not least the intrinsic spin angular momentum are ensured by the conservation of the number of ekons. The only way that the numbers can change are by unique cancellation of an ekon vanishing with another of opposite chirality or by new net generation as postulated in answering the question "What is Time". If the time scale of these transfers are too short then the further effort required to push ekons would be dramatically large. It is now postulated that this limit is in the speed of light as used presently. In the case of electro-magnetic waves this change from kinetic to potential and back takes place as a result of electric potential and current including displacement current giving raise to alternate electric and magnetic fields and it is well known that this speed is the velocity of light as defined through the electric permitivity and magnetic permeability of conventional physics.

The energy concept used currently involves the principles of conservation and linearity and energy gets related to mass and speed of light squared. This enables us to switch from the conventional units mass and velocities to ekons that is combination of both space and momentum with an assigned value having effective mass extremely small.

Mass has been determined always by a ratio like e/m, by measuring the radius of curvature of an orbit in a cloud chamber / spark chamber or by measuring the synchrotron frequencies and radius of the orbit. Direct measurement of the mass of fundamental particles without involving the electric charge is not available. Hence negative mass is interpreted as a change of charge. The same is carried over in to the so called standard model involving short range forces and electric charge is even allowed fractional values.

In a thought provoking article T.Padmanabhan has listed a number of issues. In the model of the universe proposed in this paper and the estimate of the link between the gravitational constant and the critical density of the universe there is full scope for reconciliation and new ideas.
   a. The link between high energy and low energy regimes is smooth. The small scale though small is still finite and does not lead to divergences.
   b. "Space" itself is a dynamic entity which is being created and
      Time is treated as a parameter.
   c. Finite bounds – there is nothing like being unbounded either from below or above. There is no need to deal with infinities and adopt renomalization techniques . There are no coupling constants needing to be renomalized. Gravity will certainly prove to be self destructive but there is no reason to presume that signals cannot propagate . There is no need to think of a negative time. The general impression that the space time structure gets modified at the Planck length is not valid. Rather the size contemplated is very small by several orders of magnitude.
   d. The gravitational constant $G_N$ will be the function of time and there is no reason to believe that it is constant which is the beginning of the universe. The scale of $G_N$ will adjust itself to the number of entities.
   e. The uncertainty principle has been taken care of right at the beginning . There is nothing like infinite curvature .
   f. Quantum field theory by and large have used time essentially as a parameter and the results of the special theory of relativity in converting mass to energy and back . In the considered opinion of Max Jammer the mass and energy relation through the velocity of light is still an open question and this needs separate treatment for which a frame work is available in the present approach.
   g. The ten functions $g_{ik}$ used in general relativity are an overkill.
   h. Langrangian formulation has been eschewed
      Hamiltonian formulation is more appropriate.
   i. Flatness follows effortlessly

There is nothing like infinite singularity. As matter collapses newer particles/radiation come in to account and excess energy can escape as radiation. There will be strong repulsion between the elementary entities. The result may be effectively an ordered arrangement. The effective gravitational "constant" would be small. Also there can be build up of the angular momentum. There is no need to further separate space and time. Time is only a parameter.

The conclusion of T.Padmanaban that classical general relativity is fundamentally wrong is perhaps an over statement. There is no need to bring in an extraneous observer and think of time in the observers clock. The present formulation does provide for a smooth transition to very strong curvatures. There is no need to specially invoke the slicing of space time. Already time is used as a parameter rather than another coordinate at all scales of relevance both big and small.

It is best to conclude with a sobering remark by R.C.Tolmon[8]. "It is appropriate to approach the problems of cosmology with feelings of respect for their importance, of awe for their vastness, and of the exultation for the temerity of the human mind in attempting to solve them. They must be treated however, by the detailed, critical and dispassionate methods of the Scientist"

**Acknowledgments**

For computer aided prime number searches, help from Prof. K. Ramachandra (TIFR/IISc), Prof. C.E.Veni Madhavan (IISc., Bangalore) and Mr.Vasudeva Rao (IISc., Bangalore) is gratefully acknowledged.

There is a wealth of observational data and simulation studies reported and careful modeling of the multipole structure not only in the optical region of the sky spectrum including polarization signatures but also in the noise power spectra in the radio frequency range. These offer good scope for the constraining the magnitude of various parameters.

This work would not have been possible but for the immense data collected over several decades by very large number of meticulous and unbiased observers, who are seldom mentioned while only a few hog the limelight.
May their tribe increase. My thanks to them all.

# Schematic diagram:
# Fermi-Dirac distribution of Ekons in the Universe

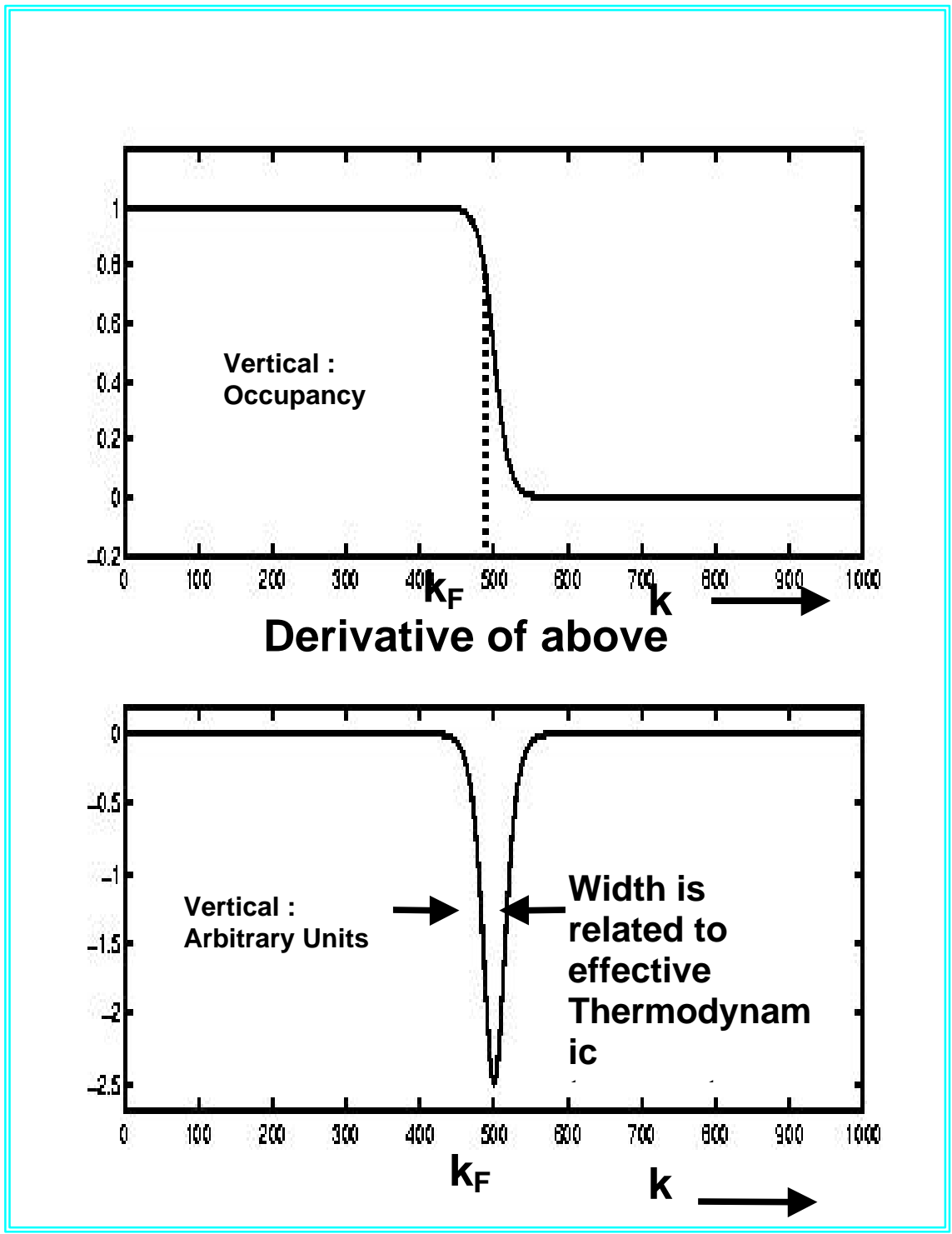

Fig. 1

**Numerical Estimates (I)** : $N_W$ number of Ekons in the universe

Radius of the universe $R_U = 9.2504 \times 10^{21}$ m
Volume of the universe = $3.315661182 \times 10^{66}$ m$^3$
Spatial part of one Ekon $(h/2)^{3/2} = 3.8288 \times 10^{-51}$ m$^3$
$N_W$ (Wheeler No.) = $8.6598 \times 10^{117}$
Number density of Ekons =
Effective Mass per Ekon $\approx$ Critical density $\div$ Number/m$^3$ = $1.8791 \times 10^{-26}$
1 year (sidereal) = $3.155815 \times 10^7$ seconds
Age of the universe (Notional) = $10 \times 10^9$ years = $3.155815 \times 10^{17}$ seconds
$[1/2(h/2\pi)]^{3/2}$ is the spatial fraction of one EKON
Volume of the universe $4\pi/3 (R_u)^3 = 3.315661182 \times 10^{66}$ m$^3$

$[1/2 \cdot (h/2\pi)]^{3/2} = 3.828858 \times 10^{-51}$
Number of Ekons = $8.865979 \times 10^{116}$

*Remarks : The latest estimate of the expansion rate of the universe with the Hubble space Telescope is reported to be
$H_o = 60 \pm 3$ kms$^{-1}$ Mpc$^{-1}$ with an rms uncertainty has been estimated to be $\pm$ 7kms$^{-1}$ Mpc$^{-1}$
An Ekon has a "volume" of $[(h/2\pi)/2]^3$. Its dimensions are momentum$^3$ x length$^3$
Position and momentum are to be treated on an equal footing. When taking the square root it is important to introduce quantities to neutralize the dimensions. Currently unit of length is the meter and unit of orbital angular momentum is $h/2\pi$ and of spin angular momentum $[(h/2\pi)/2]$.

## $G_N$ calculated on the basis of nominal age of $10^{10}$ years is as follows
$G_N = 3H_o^2 / \rho_{crit} = 3 \times 10^{-34} / (3.8563)^2 \times 1.8791 \times 10^{-26} = 5.71 \times 10^{-11}$
Ratio of $G_N$ (calculated) / $G_N$ (Known) = 6.7259 / 5.7135 = 17.5% higher
Which immediately suggests that Hubble parameter should be at least 16% smaller
Present values quoted around 30% lower in view of the crude approximation we have used regarding the size of the universe & observed orison distance the agreement is significant.

## Numerical Estimates (II) : Calculation of effective mass of an ekon

Notations as follows :

Critical density $\equiv r_{crit}$ ; Number density of ekons $\equiv n_{\hat{I}}$ (unit volume m$^3$)

Mass of one ekon $\equiv m_{\hat{I}}$ , Mass of an electron $\equiv m_e$

Number of ekons per proton $\equiv n_p$, Proton mass $\equiv m_p$,

_______________________________________________________________

$\rho_{crit}$ = 1.8791 x 10$^{-26}$ kg/m$^3$ ; $n_{\hat{I}}$ = 1 / [(h/2$\pi$ )/2]$^{3/2}$ = 2.611744807 x 10$^{51}$ /m$^3$

$m_e$ = 9.1093897(54)x10$^{-31}$kg, $m_p$ = 1.672623(10)x10$^{-27}$kg,

$m_p / m_e$ = 1836.152701 (37)

_______________________________________________________________

Effective mass of one ekon as got from critical density (kg/m$^3$) and the number of ekons / m$^3$ from the volume of the universe

$m_{\hat{I}}$ =1.8791 x 10$^{-26}$/2.6117 x 10$^{51}$ = 0.71949(305)x10$^{-77}$ kg

$m_e / m_{\hat{I}}$ =9.1093897 (54) x 10$^{-31}$ / 0.71949305 x 10$^{-77}$

Ratio No. of ekons per electron = 1.266608 x 10$^{47}$

Find nearest primes and compare the ratio

Find primes nearest 1266608 x 10$^{41}$
        Proton - 2104112 x 10$^{44}$

m_electron  12820894 * 10 ^ 40 + 1
m _proton    2354112  * 10 ^ 44 + 1
m _proton / m_electron = 1836.152767 ( Calculated value)  *
m _proton / m_electron = 1836.152701 (Measured value)

\*  It must be noted that in arriving  at this ratio two adjustments have been made viz i)the size of the universe as got from the Hubbles parameter and  ii) effective critical density as known now.  Another method of constraining  would be to   use the known ratio itself ( which must be rational) to the available accuracy and calculate the nearest prime numbers.

Prime number search for 1266608 x 10$^{41}$ was done with a constraint regarding the number of ekons per proton $p_{\hat{I}} / n_{\hat{I}}$ ratio taken as the well determined proton/electron mass ratio as given in standard literature as 1836.152701(37) .  Values are as given in excellent and useful collection of the physical constants and other data, see " Physical formulas by graham Woan, CUP-2000.

A significant fact to be noticed is that even though both the spin of  electron and proton are [½(h/2$\pi$)] their magnetic moments are vastly different suggesting much slower movement of electric charges (Bohr magneton versus nuclear magneton).  In the standard model the almost point like structure of the ultimate structure  is indicated including the fractional electric charges.

**Numerical Estimates III : Example for Electromagnetic Radiation**

Let us consider the CMBR spectrum. It fits the perfect Black Body spectrum at radiation temperature Longair M.S[15]

$2.728 \pm .002$k with a peak at 160 GHz

Frequency at the peak as read form the graph $\rightarrow$ 160 GHz

Corresponding wave length $\lambda$ as read off from the graph
Corresponding wave length $\lambda$ calculated from $c = \nu \lambda$

$\nu = 160 \times 10^9$, $c = 2.99792 \times 10^8$ ms$^{-1}$

$\lambda = c/\nu = 2.99792 \times 10^8 / 16 \times 10^9 = .018737 \times 10^{-1}$
$\qquad\qquad\qquad\qquad\qquad\qquad\quad = .0018737$ m

$\qquad\qquad\qquad\qquad \lambda = 1.8737$mm

corresponding momentum $h\nu/c = h/\lambda = 6.6260755 \times 10^{-34}/1.873 \times 10^{-3}$
$\qquad\qquad\qquad\qquad\qquad\qquad = 3.537680459 \times 10^{31} \times 10^{-34}$

Corresponding energy $\qquad = h\nu = 6.6726 \times 10^{-34}$ joules sec. $\times 160 \times 10^9$
of one photon $\qquad\qquad\qquad = 1067.616 \times 10^{-34} \times 10^9$
at the peak $\qquad\qquad\qquad\quad = 1.067 \times 10^3 \times 10^{-34} \times 10^9$
$\qquad\qquad\qquad\qquad\qquad = 1.067 \times 10^{-22}$ Joules

relate this energy to the Thermal energy $3/2$ kT

T $\qquad\qquad = 2.728\ ^0$ kelvin
$k_B \qquad\qquad = 1.3806 \times 10^{-23}$ Joules k$^{-1}$
$3/2\ k_B$ T $\qquad = 5.6494152 \times 10^{-23}$ Joules
Ratio $\qquad\quad = 1.067 \times 10^{-22} / 5.649 \times 10^{-23} = 0.188883 \times 10^1$
$\qquad\qquad\quad = 1.888 \approx 2$ two helicities for the photon

For radiation it is momentum that has to be compared.

# Essential Bibliography with notes and short quotes

This result is highly relevant to the discussion and estimate of the Newtonian gravitational Constant $G_N$ .

" Quantum nonlocality differs from nonlocal action entailed by classical nonrelativistic theories (such as Newtonian gravity, electrostatics, heat diffusion) in that it implies action at a distance that does not diminish in strength with increasing distance. Basically quantum nonlocality is kinematic in nature and pertains to cases where correlation properties embodied in the nonfactorisable quantum mechanical wave functions (the kinematic component of the theory) are not fully reproducible by a realist theory satisfying the locality condition. This incompatibility is experimentally verifiable and it can be theoretically demonstrated by using arguments that do not depend on the way a measurement process is described"
" Studies to this end not only deepen our understanding of the "vulnerable" areas of quantum mechanics but also reveal hitherto unexplored facts of the theory. For instance, the empirically relevant manifestations of quantum entanglement that have been discovered related to quantum nonlocality have of late acquired considerable significance in view of their applications in the context of cryptography, quantum communications and quantum computing. Novel approaches have been developed to gain fresh insights into the quantum measurement problem which will make it empirically investigable, like biomolecular systems being used as mesoscopic quantum measuring devices. Ingenious new models of quantum mechanics have also been proposed which have the potentiality of yielding predictions not contained within the standard framework".
The author would like to emphasize that the finite growing universe model based on Ekons are able to fully address all these problems in a simple and easily understandable way.

An article by Witten : "Through the Clouds - Actually there is one major approach to the problem that I have not yet mentioned. The question of why the weak scale is so small is really a modern version of the problem that Dirac in the 1930's called the problem of the large numbers. His first "large number" was the ratio of the Planck mass to the proton mass, which is $N_1 \sim 10^{19}$.

Dirac's idea was to relate this large number to another conspicuous large number in physics, namely the age of the universe in nuclear time units; this is $N_2 \sim 10^{41}$. Dirac proposed that the gravitational constant G changes in time to maintain a rough relation $N_2 \sim N_1^2$.

Unfortunately , Dirac's lovely idea, one of the most beautiful suggestions that has been made about this problem, seems to be ruled out by experiment - most directly, by measurements of G/$G_N$ (the rate of change time of the gravitational constant) made possible by a radio transponder landed on Mars by a Viking space craft.

Whether or not Dirac's idea can somehow be revived, the large number problem that he singled out is one off the keys to astronomy and life. The existence of complex structures in the universe, containing many elementary particles, existing for long periods, capable of storing information and giving birth to life, certainly depends among other things on the solution to the problem of large numbers.

For instance, the theory of star formation shows in its most simple considerations that the number of elementary particles in a star is proportional to $N_1^3$. This is why the mass of the Sun is about $10^{33}$ grams. Had nature not solved the large number problem, there would be no astronomy as we know it , and certainly no life.

Finally, it will be seen as an era in which the geometrical tools at our disposal lag behind what we need to properly appreciate the insights glimpsed through the clouds, and in which the <u>search for geometric understanding is thus likely to be one of the great engines of progress"</u>.

Sivaram C, Current Science, *Some implications of quantum gravity and string theory for everyday physics* vol.79, Pages 413-420, 2000 : While conceding Quantum gravity is still an enigmat , the author has brought together a number of numerical estimates and other links to various current theories.

Somnath Bharadwaj *Proceedings of the workshop and Sayankar on cosmology* : observations confront theories, special issue of Pramana- Journal of physics vol.**5 3**, **6**, 1999

Straumann .N, *The Mystery of the cosmic vacuum energy density and the accelerated expansion of the universe,* Eur. J.Phys.(419-427)*,* IOP Publishing Ltd. 1999 : For a short and clear account of the problems associated with the cosmological constants and related matters.

Stefano Borgani & Luigi Guzzo, X-ray clusters of galaxies as tracers of structure in the Universe, Nature, vol409, page 39-46, 2001